\documentclass[twocolumn]{aastex62}
\usepackage{natbib}
\usepackage{color}
\usepackage{graphicx}
\bibpunct{(}{)}{;}{a}{}{,}

\received{January 1, 2018}
\revised{January 7, 2018}
\accepted{\today}
\submitjournal{ApJ}

\shorttitle{Gas-phase metallicity of an ultra-diffuse galaxy}
\shortauthors{S\'anchez Almeida et al.}

\begin{document}

\title{A headless tadpole galaxy: the high gas-phase metallicity of the ultra-diffuse galaxy UGC\,2162}

\correspondingauthor{J. S\'anchez Almeida}
\email{jos@iac.es}

\author[0000-0003-1123-6003]{J. S\'anchez~Almeida}
\affiliation{Instituto de Astrof\'\i sica de Canarias\\
La Laguna, Tenerife, Spain}
\affiliation{Departamento de Astrof\'\i sica,\\
Universidad de La Laguna, Tenerife, Spain}

\author[0000-0001-7859-699X]{A. Olmo-Garc\'\i a}
\affiliation{Instituto de Astrof\'\i sica de Canarias\\
La Laguna, Tenerife, Spain}
\affiliation{Departamento de Astrof\'\i sica,\\
Universidad de La Laguna, Tenerife, Spain}

\author[0000-0002-1723-6330]{B.~G. Elmegreen}
\affiliation{ IBM Research Division, T.J. Watson Research Center\\
1101 Kitchawan Road, Yorktown Heights\\
NY 10598, USA}

\author[0000-0002-1392-3520]{D.~M. Elmegreen}
\affiliation{Department of Physics \& Astronomy, Vassar College\\
 Poughkeepsie, NY 12604, USA}

\author[0000-0001-7958-4536]{M.~Filho}
\affiliation{Center for Astrophysics and Gravitation - CENTRA/SIM\\
Departamento de F\'\i sica, Instituto Superior T\'ecnico, Universidade de Lisboa\\ 
Av. Rovisco Pais 1, P-1049-001 Lisbon, Portugal}
\affiliation{Departamento de Engenharia F\'\i sica, Universidade do Porto\\
Faculdade de Engenharia, Universidade do Porto\\
Rua Dr. Roberto Frias, s/n, P-4200-465, Oporto, Portugal
}

\author[0000-0001-8876-4563]{C.~Mu\~noz-Tu\~n\'on}
\affiliation{Instituto de Astrof\'\i sica de Canarias\\
La Laguna, Tenerife, Spain}
\affiliation{Departamento de Astrof\'\i sica,\\
Universidad de La Laguna, Tenerife, Spain}

\author[0000-0003-3985-4882]{E.~P\'erez-Montero}
\affiliation{Instituto de Astrof\'\i sica de Andaluc\'\i a -- CSIC,\\
Granada, Spain Granada}

\author[0000-0002-3849-3467]{J.~Rom\'an}
\affiliation{Instituto de Astrof\'\i sica de Canarias\\
La Laguna, Tenerife, Spain}
\affiliation{Departamento de Astrof\'\i sica,\\
Universidad de La Laguna, Tenerife, Spain}

%
%




\begin{abstract}
The cosmological numerical simulations tell us that accretion of external metal-poor gas drives  star-formation (SF) in galaxy disks. One the best pieces of observational evidence supporting this prediction is the existence of low metallicity star-forming regions in relatively high metallicity host galaxies. The SF is thought to be fed by metal-poor gas recently accreted. Since the gas accretion is stochastic, there should be galaxies with all the properties of a host but without the low metallicity starburst. These galaxies have not been identified yet. The exception may be UGC\,2162, a nearby ultra-diffuse galaxy (UDG) which combines low surface brightness and relatively high metallicity. We confirm the high metallicity of UGC\,2162 ($12+\log({\rm O/H})=8.52^{+0.27}_{-0.24}$) using spectra taken with the 10-m GTC telescope.
UGC\,2162 has the stellar mass, metallicity, and star-formation rate (SFR) surface density expected for a host galaxy in between outbursts.  This fact suggests a physical connection between some UDGs and metal-poor galaxies, which may be the same type of object in a different phase of the SF cycle. UGC\,2162 is a high-metallicity outlier of the mass-metallicity relation, a property shared by the few UDGs with known gas-phase metallicity.  
%
%
\end{abstract}



\section{Scientific rationale} \label{sec:intro}

The current cosmological numerical simulations predict that accretion of metal-poor gas from the cosmic web drives the star-formation  in galaxy disks \citep[e.g.,][]{2008A&ARv..15..189S,2009Natur.457..451D,2012RAA....12..917S,2014A&ARv..22...71S}. One of the best pieces of observational evidence in favor of this external feeding is the existence of low metallicity star-forming regions in relatively high metallicity host galaxies \citep[][]{2013ApJ...767...74S,2015ApJ...810L..15S}. The main low-metallicity starburst tends to be off-centered so the galaxies commonly have a cometary or tadpole morphology \citep{2008A&A...491..113P,2011ApJ...743...77M}. In terms of their average metallicity, these galaxies are often extremely metal poor (XMP), with a  light-weighted mean gas-phase metallicity smaller than a tenth of the value in the Sun ($Z_g < Z_\odot/10$).  Their starbursts are thought to be fed by metal-poor gas recently accreted by the  galaxy, a process that has been recently modeled by \citet[][]{2014MNRAS.442.1830V} and \citet[][]{2016MNRAS.457.2605C}. 
The galaxies hosting the XMP starburst are low surface brightness galaxies with a metallicity around half the solar metallicity \citep[$Z_g \sim Z_\odot/2$;][]{2015ApJ...810L..15S}. We will call them {\em host galaxies}. Since the gas accretion process is stochastic, there should be galaxies with the properties of the host galaxies, i.e., relatively high metallicity for their mass, low surface brightness, and without the bright low metallicity star-forming region (i.e., {\em headless tadpole galaxies}).
%
%
However, such galaxies have not been identified before. Low surface brightness galaxies are also faint galaxies \citep[e.g.,][]{1999ASPC..170..169S} which, according to the luminosity--metallicity relation \citep[e.g.,][]{2012ApJ...754...98B}, are metal poor.  Actually, a few known low surface brightness dwarf galaxies with mean metallicity determined through the direct-method (DM; details in Sect.~\ref{subsec:metallicity}) seem to be XMP 
(e.g., Leo~P by \citeauthor{2013AJ....146....3S}~\citeyear{2013AJ....146....3S};
 Leoncino Dwarf by \citeauthor{2016ApJ...822..108H}~\citeyear{2016ApJ...822..108H}; and
Little Cub by \citeauthor{2017ApJ...845L..22H}~\citeyear{2017ApJ...845L..22H}). 

\begin{figure}[ht!]
\plotone{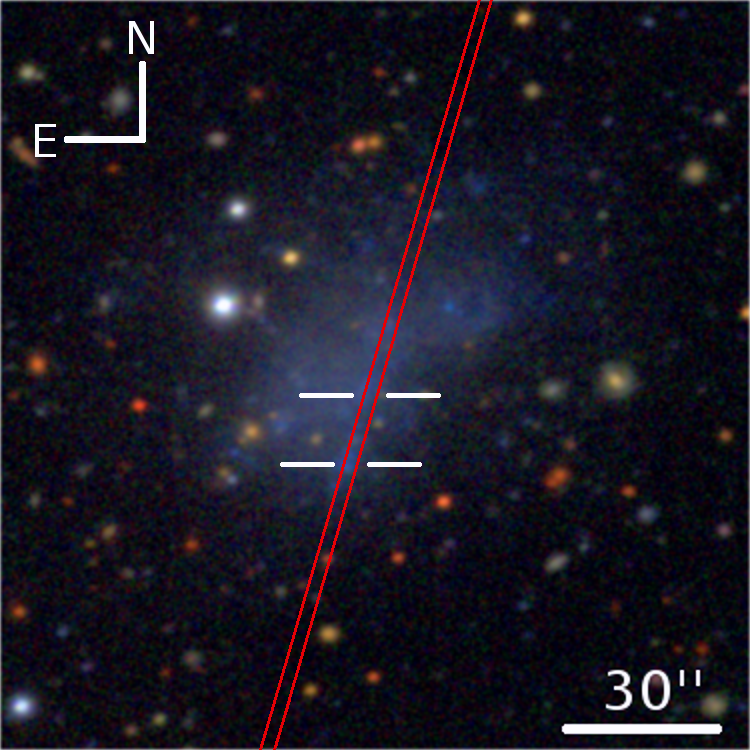}
\caption{Composite image of UGC~2162 in $g$, $r$, and $i$ from the IAC SDSS Stripe 82 legacy project \citep{2016MNRAS.456.1359F}.  The white  ticks indicate the location of the two star-forming regions mentioned in the main text, with their spectra shown in Figs.~\ref{fig:gtc_spect} and \ref{fig:spect}. The orientation of the spectrograph slit during GTC observations is shown in red. A 30\,arcsec  scale has been included for reference. North (N) and east (E) are also indicated.
\label{fig:fig1}
}
\end{figure}

UGC~2162 may be an exception compared to these low surface brightness dwarf galaxies with known metallicity, and so, it may be the sought-after headless tadpole galaxy that provides support to the whole picture.  It may be one of the expected host galaxies, without a metal-poor starburst. It has been recently identified as a nearby low-mass UDG\footnote{The term UDG (Ultra Diffuse Galaxy) was recently introduced by \citet{2015ApJ...798L..45V}, and it quickly became widely used in the literature despite the fact it describes galaxies which were already known as low surface brightness dwarf ellipticals or dwarf irregulars,  depending on whether they reside in clusters or the field  \citep[e.g.,][]{1997ARA&A..35..267I,1997AJ....114..635D,2018RNAAS...2a..43C}. We have decided to stick to the now popular terminology, acknowledging that the word UDG could be replaced throughout the text with {\em low surface brightness dwarf  irregular} without altering the content of our paper.} by \citet{2017ApJ...836..191T},  with $M_\star\simeq 2\times 10^7 M_\odot, M_{\rm HI}/M_\star\simeq 10,$ and $M_{dyn}/M_\star\simeq 200$ ($M_\star$, $M_{\rm HI}$, and $M_{dym}$ stand for the stellar mass, the HI mass, and the dynamical mass, respectively). With an effective radius of 1.7 kpc, its $g$-band central surface brightness is 24.4 mag~arcsec$^{-2}$. The galaxy is irregular with several bluish star-forming knots (Fig.~\ref{fig:fig1}). UGC~2162 belongs to the group of M77, even though it is quite far from the central galaxy \citep[293$\pm 40$ kpc projected distance;][]{2017ApJ...836..191T}.
One of these knots has a spectrum in the Sloan Digital Sky Survey Data \citep[SDSS-DR12,][]{2015ApJS..219...12A}. Using the line ratio N2 ([NII]$\lambda$6583/H$\alpha$) derived from the SDSS spectrum,  and the calibration by \citet{2004MNRAS.348L..59P}, \citet{2017ApJ...836..191T} estimate an oxygen abundance for  the star-forming ionized gas of $12+\log({\rm O/H})=8.22\pm 0.07$, which corresponds to 1/3 of the solar abundance\footnote{With $12+\log({\rm O/H})_\odot = 8.69$, from \citet{2009ARA&A..47..481A}.}. If this metallicity estimate is correct, the gas of UGC\,2162 is of high metallicity for its mass and magnitude, in the vein expected for the galaxies that, when hosting a metal-poor starburst, give rise to the observed XMP galaxies\footnote{Here and throughout the paper, we use the term {\em XMP galaxy} to denote a host galaxy plus one or a few metal-poor starbursts which outshine the emission line spectrum of the host, so that the light-weighted average metallicity of the combined system is smaller than a tenth of the solar metallicity.}.
\begin{figure}[ht!]
\includegraphics[width=10cm, bb= 70 50 560 400]{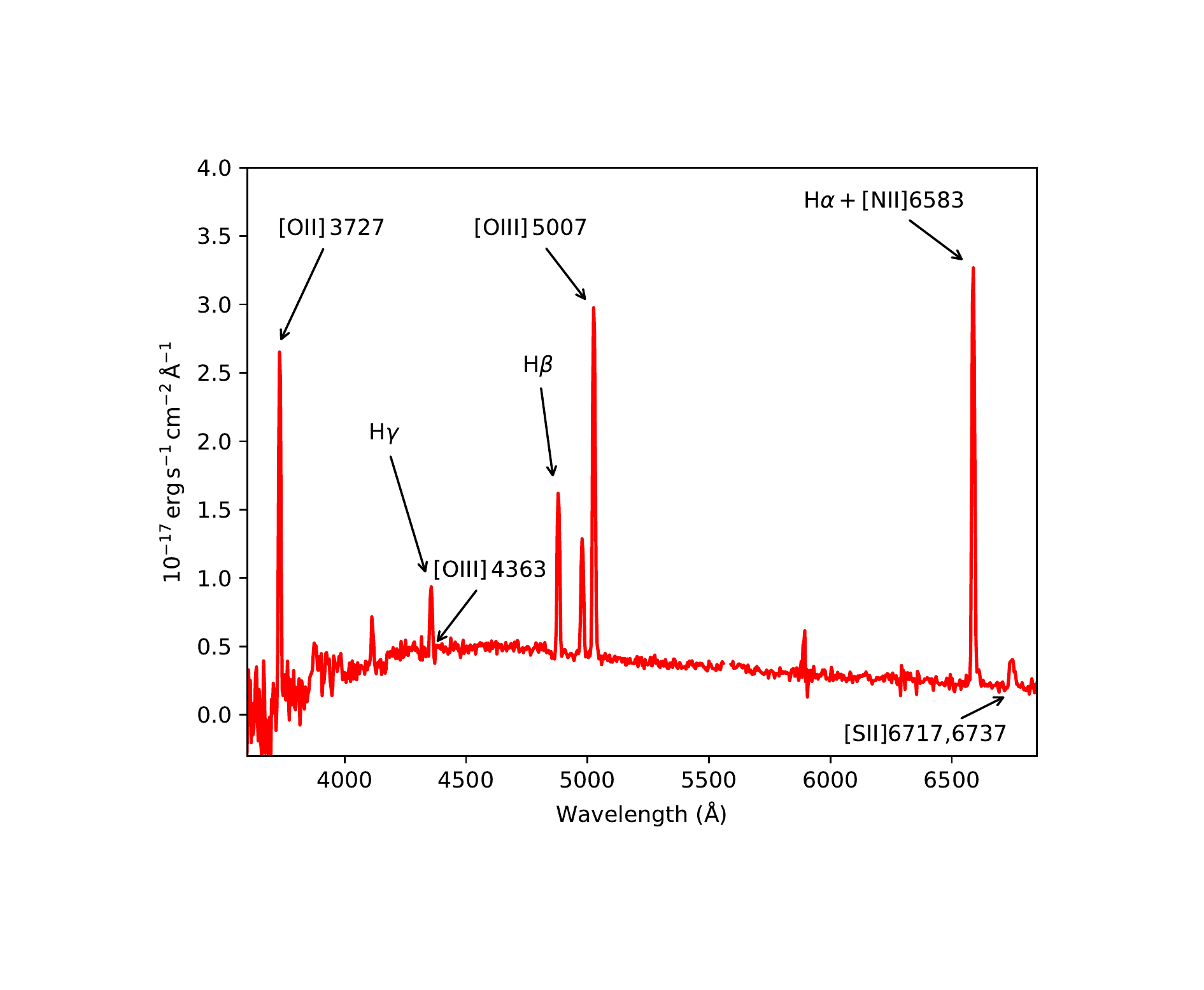}
\caption{
GTC spectrum corresponding to the brightest knot of UGC~2162 (the southmost in Fig.~\ref{fig:fig1}). The position of the main emission lines used to estimate the metallicity are labeled, including [OIII]$\lambda$4363, even though is below the detection threshold. H$\gamma$ is also labeled for clarity to avoid confusion with [OIII]$\lambda$4363.
\label{fig:gtc_spect}
}
\end{figure}

The reported high metallicity of  UGC\,2162 rests only on the low signal-to-noise ratio (S/N) SDSS spectrum,  and on the estimate of its metallicity using N2. Due to the importance of the gas-phase metallicity of this particular galaxy, we obtained long-slit spectra of  UGC\,2162 with the 10-m GTC telescope, with enough wavelength coverage and S/N to carry out an independent robust determination of the gas-phase metallicity. The analysis of these spectra confirms the high metallicity of the star-forming gas in UGC\,2162 (Sect.~\ref{sec:obs}). The purpose of this paper is to present this determination, and to show that UGC\,2162 shares many properties in common with those expected for the host galaxies having XMP star-forming regions  (Sect.~\ref{sec:results}). There seem to be a few other objects in the literature with the characteristics of UGC\,2162. Discussions are included in Sect.~\ref{sec:conclusions}.

\section{Observations and metallicity determination}\label{sec:obs} %

\subsection{Observations}

We obtained long-slit spectra of UGC\,2162 integrating for 2 hours with the instrument OSIRIS at the 10-m GTC telescope\footnote{{\tt http://www.gtc.iac.es/instruments/osiris/osiris.php}}. The 1~arcsec wide slit was placed as shown in Fig.~\ref{fig:fig1}, inclined with respect to the major axis so as to cover the two main star-forming regions of the galaxy and the diffuse emission around them. The resulting visible spectra span from 3600 to 7200\,\AA\ with a spectral resolution around 550, thus covering the wavelength range containing [OII]$\lambda$3727 and the temperature sensitive line [OIII]$\lambda$4363. The raw data were reduced following the usual procedure which includes correction for bias and flat-field, and flux and wavelength calibration. We employ  
PyRAF\footnote{PyRAF is a product of the Space Telescope Science Institute, which is operated by AURA for NASA {\tt http://www.stsci.edu/institute/software\_hardware/pyraf/}}
for the task. Figure~\ref{fig:gtc_spect} shows one of the resulting spectra, corresponding to the brightest knot of the galaxy (the southmost in Fig.~\ref{fig:fig1}). The slit was not oriented along the parallactic angle, however, differential refraction is not an issue since the airmass of observation is low ($<\,1.2$), and the residual effect is corrected for by the flux calibration. 


\subsection{Metallicity determination}\label{subsec:metallicity}
\begin{figure}[ht!]
\includegraphics[width=10cm, bb= 70 50 560 450]{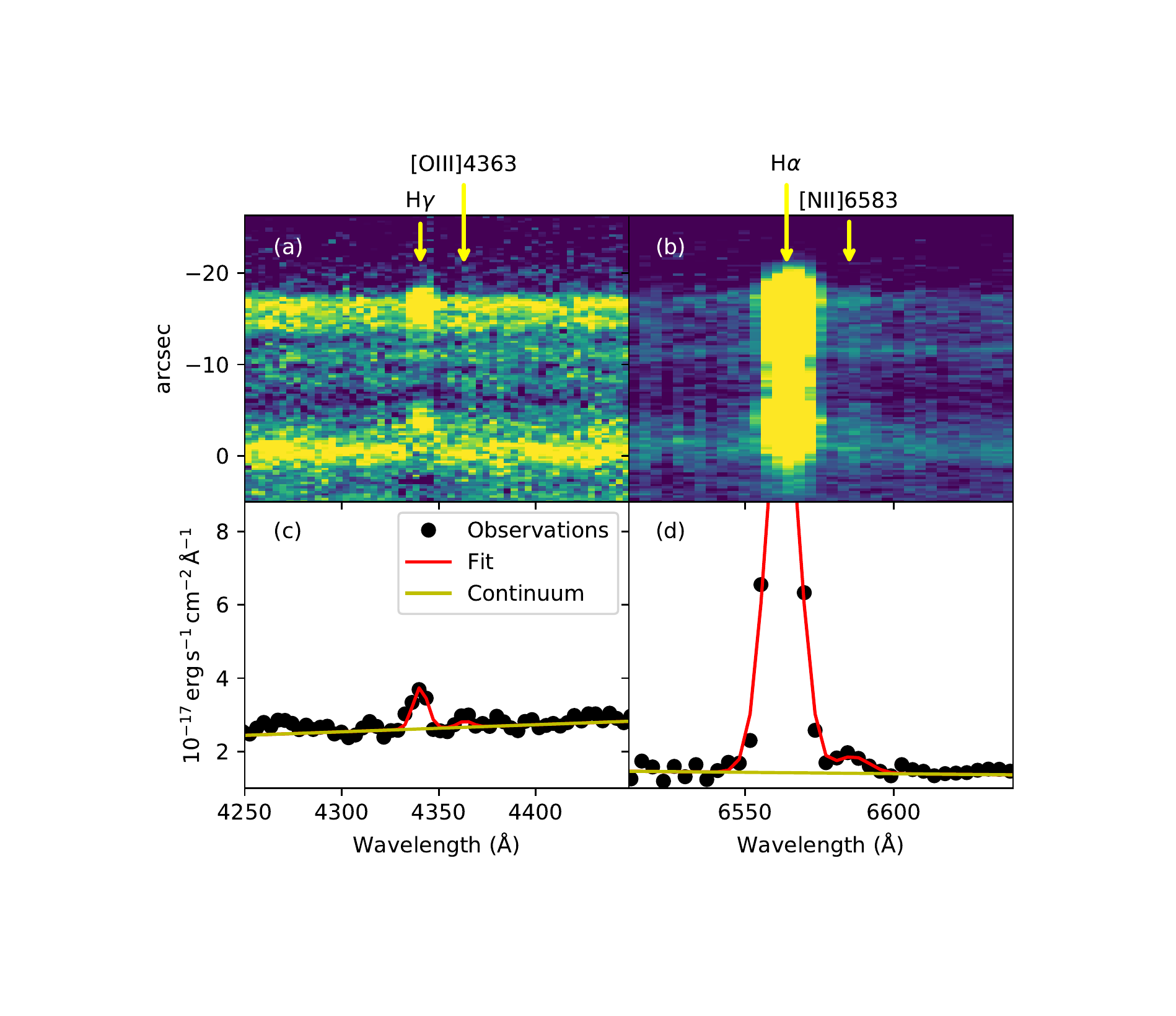}
\caption{(a) Image with the part of the 2D spectrum which should contain [OIII]$\lambda$4363. The horizontal and the vertical directions correspond to wavelength and position along the slit, respectively. The color palette goes from dark blue to yellow from low to high signal.
H$\gamma$,  at  4340 \AA\, shows up clearly. However, any possible signal of  [OIII]$\lambda$4363 is buried within the noise level.
(b) Same as (a) for the spectral region around H$\alpha$. The faint [NII]$\lambda$6583 is barely visible in the red wing of H$\alpha$.
(c) Fit with two Gaussians and a continuum to the spectral region around [OIII]$\lambda$4363. The observed spatially integrated spectrum of the galaxy is presented as solid symbols whereas the red and the yellow lines stand for the full fit and the continuum, respectively.
(d) Same as (c) for the spectral region around H$\alpha$. [NII]$\lambda$6583 stands out above the noise level.
\label{fig:spect}
}
\end{figure}

The DM is the method of reference to measure metallicities when all the required emission lines are available \citep[e.g.,][]{2004cmpe.conf..115S,2017PASP..129d3001P}. It obtains the physical properties of the emitting gas from the same spectrum used to determine the metallicity, minimizing the model dependence of the result. In the case of the oxygen abundance, the emission-line ratio [OIII]$\lambda$4363/[OIII]$\lambda$5007 is needed to determine the electron temperature of the nebulae and so to determine the oxygen abundance using the DM \citep[e.g.,][]{1974agn..book.....O,2017PASP..129d3001P}. As it happened with the SDSS spectrum analyzed by \citet[][]{2017ApJ...836..191T}, [OIII]$\lambda$4363 does not show up above the noise level even in the GTC spectra. The 2D spectrum around  [OIII]$\lambda$4363 is shown in Fig.~\ref{fig:spect}a. 
[OIII]$\lambda$4363 is prominent in low metallicity HII regions \citep[e.g.,][]{2015ApJ...805...45L,2016ApJ...819..110S} and so the non-detection already suggests the high metallicity of the gas in the galaxy.  On the other hand, the non-detection implies that we cannot use the DM to infer the gas-phase metallicity. We employ the code HII-CHI-mistry (HCm) to determine O/H. HCm compares the observed emission line fluxes of a selected number of lines with a grid of photoionization models, finding the best fit through a least-squares minimization algorithm \citep[][]{2014MNRAS.441.2663P}. In principle, HCm employs the fluxes of [OII]$\lambda$3727, [OIII]$\lambda$4363, [OIII]$\lambda$5007, [NII]$\lambda$6583, and [SII]$\lambda\lambda$6717,6737,  relative to H$\beta$ and corrected for reddening. However, HCm was chosen because it can be used without [OIII]$\lambda$4363, and it is robust in the sense of being equivalent to the direct method when [OIII]$\lambda$4363 is available \citep{2014MNRAS.441.2663P,2016ApJ...819..110S,2017A&A...601A..95C}. The fluxes of the required emission lines were obtained fitting a Gaussian plus a linear continuum to each emission line. In the case of two overlapping lines, the two Gaussians were fitted simultaneously. Examples are given in Figs.~\ref{fig:spect}c and \ref{fig:spect}d. Dust extinction was considered and corrected for following the usual approach of assuming a Milky Way extinction law \citep{1989ApJ...345..245C}, with the ratio between H$\alpha$ and H$\beta$ given by the case B recombination at  high electron temperatures in HII regions \citep[i.e., 2.76;][]{1974agn..book.....O}.  The flux ratios thus obtained are collected in Table~\ref{tab:fluxes}, with the corresponding oxygen abundances given in Table~\ref{tab:logoh}, column\,1. 
The four rows correspond to four different spatial averages of the observed long-slit spectrum: the full galaxy, two bright points, and the diffuse component. In terms of the distances represented in Fig.~\ref{fig:spect}a, the full galaxy considers the average spectrum from $-21\arcsec$ to $+1\arcsec$, the first bright point from $-19\arcsec$\ to $-15\arcsec$, the second bright point from $-8\arcsec$\ to $-3\arcsec$, and the diffuse component from $-15\arcsec$ to $-6\arcsec$ (see also the ticks in Fig.~\ref{fig:fig1}). Thus the spectra of the two bright points and the diffuse component come from different parts of UGC\,2162, whereas the full galaxy spectrum basically averages out the three of them.  
In all four cases the metallicity is relatively high, with $12+\log({\rm O/H})$ between 8.3 and 8.7 (Table~\ref{tab:logoh}, column\,1).
This range of values is within the error bar of the measurement, as we discuss in the next paragraph.
%
%
%

\begin{deluxetable*}{lccccc}
\tablecaption{Emission line flux ratios used in the metallicity determination\label{tab:fluxes}}
\tablehead{
\colhead{Component}&
\colhead{[OII]$\lambda$3727}&
\colhead{[OIII]$\lambda$4363}&
\colhead{[OIII]$\lambda$5007}&
\colhead{[NII]$\lambda$6583}&
\colhead{[SII]$\lambda\lambda$6717,6737}
}
\startdata
Full Galaxy& $2.35\pm0.17$&$< 0.04$& $1.61\pm 0.05$& $0.17\pm 0.05$& $0.94\pm 0.05$\\
1st Bright Point&$1.91\pm 0.14$&$< 0.02$& $2.11\pm 0.07$& $0.09\pm 0.05$& $0.43\pm 0.05$\\
2nd Bright Point&$2.29\pm 0.20$& $< 0.05$& $1.50\pm 0.05$& $0.11\pm 0.05$& $0.92\pm 0.05$\\
Diffuse Emission&$2.89\pm 0.37$& $< 0.05$& $0.99\pm  0.05$& $0.13\pm 0.06$& $0.96\pm 0.05$
\enddata
\tablecomments{
The dimensionless flux ratios are relative to H$\beta$ and corrected for reddening assuming H$\alpha$/H$\beta$ = 2.76.
Error bars are statistical errors inferred from fitting the observed emission lines to Gaussian functions. 
} 
\end{deluxetable*}

%
%
\begin{deluxetable*}{lccccc}
\tablecaption{Different estimates of $12+\log({\rm O/H})$ in UGC\,2162\label{tab:logoh}}
\tablehead{
\colhead{Component}&
\colhead{HCm}&\colhead{HCm}&\colhead{Direct Method}&\colhead{N2-based}&\colhead{N2-based}\\
&\colhead{No [OIII]$\lambda$4363}&\colhead{Monte Carlo}&\colhead{Monte Carlo}&\colhead{GTC}&\colhead{SDSS}\\
&\colhead{(1)}&\colhead{(2)}&\colhead{(3)}&\colhead{(4)}&\colhead{(5)}
}
\startdata
Full Galaxy& 8.38$\pm$0.06&\dots&\dots&8.12$\pm$0.11&8.22$\pm$0.07\\
1st Bright Point&8.29$\pm$0.08&$8.52^{+0.27}_{-0.24}$&$8.22^{+0.39}_{-0.18}$&8.05$\pm$0.14&\dots\\
2nd Bright Point&8.40$\pm$0.06&\dots&\dots&8.11$\pm$0.10&\dots\\
Diffuse Emission&8.70$\pm$0.02&\dots&\dots&8.14$\pm$0.12&\dots
\enddata
\tablecomments{
(1) Values with the formal error bars provided by HCm.
(2) Monte-Carlo simulation assuming the [OIII]$\lambda$4363 flux randomly distributed with values below the observed noise. The table lists the median and the 1-$\sigma$ error bar of the resulting distribution, with the error given by the values between percentiles 15.9\,\%\ and 84.1\,\%.  
(3) Same as (2) for metallicities based on the DM.
(4) Based on the index N2, with the same calibration employed by \citet{2017ApJ...836..191T}. 
(5) As given by {\citet{2017ApJ...836..191T}}.
}
\end{deluxetable*}

HCm provides formal error bars from the difference between the observed fluxes and those provided by the photoionization models. They are rather small ($< 0.1\,$dex; see Table~\ref{tab:logoh}), and do not include the uncertainty associated with the non-detection of [OIII]$\lambda$4363. In order to be conservative and include this other source of error, we developed an alternative way to evaluate the error bars of the measured abundances. We carried out a Monte-Carlo simulation where [OIII]$\lambda$4363 was assumed to have a flux smaller than the upper limit set by the noise of the GTC spectra. Specifically, we assume [OIII]$\lambda$4363 to be drawn from a uniform distribution consistent with the noise in the observation, with values going from zero to the value set by the Gaussian function fit to the noise shown in Fig.~\ref{fig:spect}c. We drew 1000 random values, from which we compute the median and the 1-$\sigma$ error bar (i.e., the range of values in between percentiles 15.9\,\% and 84.1\,\%). The result for the brightest point is given in Table~\ref{tab:logoh}, column\,2. The measurement, along with its error bar, provides the range of values of $12+\log({\rm O/H})$ consistent with the noise level of the GTC spectra. The range is consistent with the value inferred from a single application of HCm, except that the error bars are significantly larger. We repeated the same exercise with the metallicities inferred applying the DM\footnote{We followed the step-by-step prescription described by \citet{2017PASP..129d3001P}.}. The result is given in Table~\ref{tab:logoh}, column\, 3. Once again, the result is consistent with the high metallicity inferred by \citet{2017ApJ...836..191T} from the SDSS spectrum. 
In the case of  \citet{2017ApJ...836..191T}, they used the line ratio N2 ([NII]$\lambda$6583/H$\alpha$) to estimate $12+\log({\rm O/H})$ as calibrated by \citet{2004MNRAS.348L..59P}. We have repeated the same exercise with the GTC spectra, and the metallicities inferred from N2 also agree:  compare the N2-based metallicity from the full galaxy GTC spectra, Table~\ref{tab:logoh} column\,4, with the metallicity worked out in \citet{2017ApJ...836..191T},  Table~\ref{tab:logoh} column\,5. One final comment is in order. Note that the metallicity estimated for the diffuse component using HCm differs by 0.4 dex from the metallicity of the brightest star-forming knot (Table~\ref{tab:logoh}, column\, 1). In view of the error bars inferred from the Monte-Carlo simulation, we do not consider this difference to be significant, and our observation is consistent with UGC\,2162 having a uniform metallicity.

N/O in local galaxies increases with increasing gas phase metallicity \citep[e.g.,][]{2016MNRAS.458.3466V}, and can be used to constrain O/H.  HCm also provides an estimate for N/O, which is consistent with O/H if UGC\,2162 follows the trend observed in local galaxies \citep[e.g.,][]{2016MNRAS.458.3466V}. The 1st bright point turns out to have $\log({\rm N/O})= -1.67\pm 0.03$, or $-1.52\pm 0.08$ if the Monte Carlo simulation is considered. These values correspond to $12+\log({\rm O/H}) < 8.5$ \citep[e.g., Fig.~1 in ][]{2016MNRAS.458.3466V}, and so are compatible with the other determinations of the metallicity in UGC\,2162.

\section{Results and Implications}\label{sec:results} 

\begin{figure}[ht!]
\includegraphics[width=10cm, bb= 70 50 560 450]{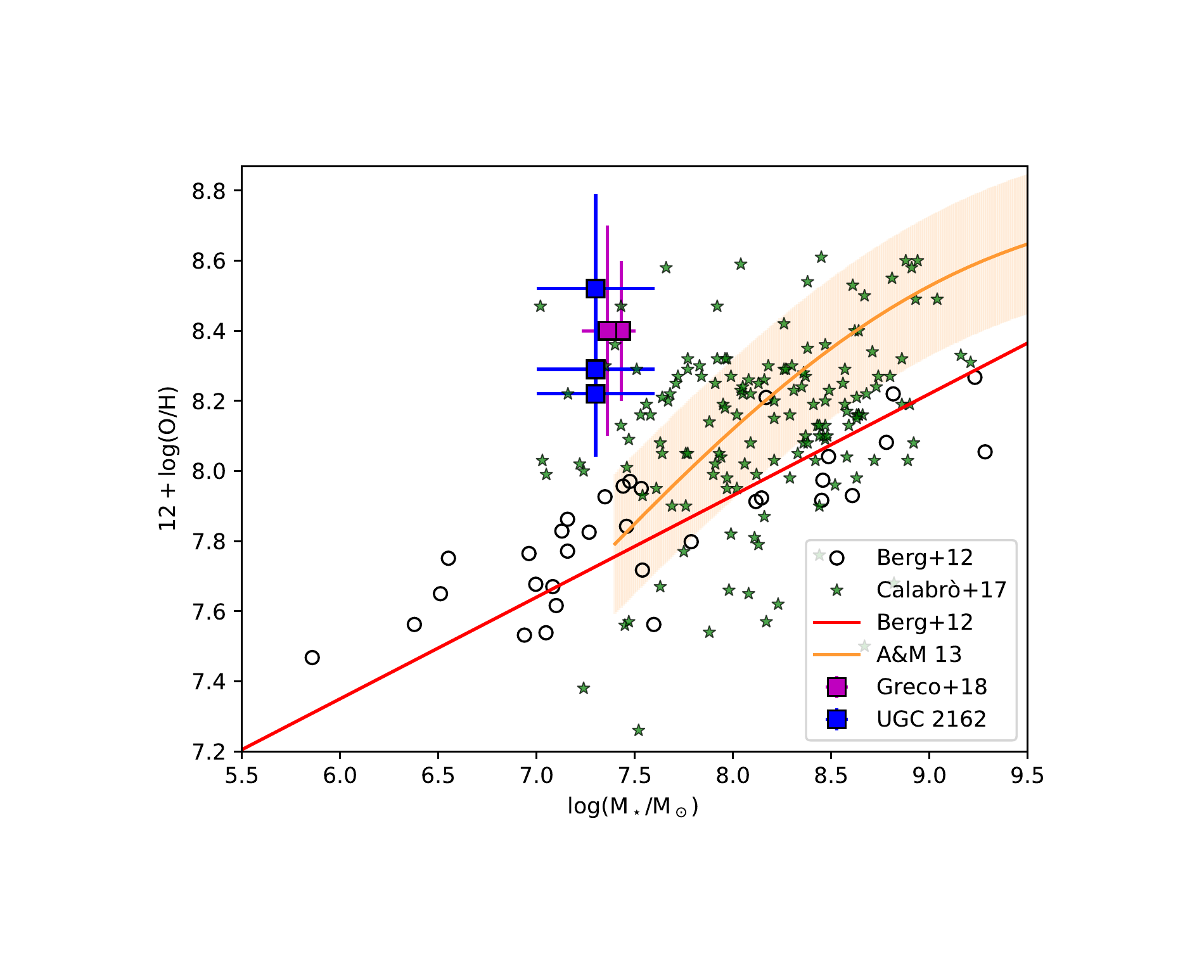}
\caption{
UGC\,2162 in the MZR (gas-phase metallicity versus $M_\star$). The blue square symbols correspond to three different gas-phase metallicities for the brightest star-forming knot in UGC\,2162 inferred from the GTC spectra (using HCm directly, HCm through a Monte Carlo simulation, and the DM; see Table~\ref{tab:logoh}).  For reference, the figure includes the MZR derived for local dwarf galaxies by \citet[][the red solid line, with the empty circles being the original data]{2012ApJ...754...98B}, as well as the relation inferred for all star-forming SDSS galaxies by \citet[][the orange solid line, with the shaded region representing the 1-$\sigma$ dispersion]{2013ApJ...765..140A}. Other objects occupying the same region as UGC\,2162 are also shown  (green stars, \citeauthor{2017A&A...601A..95C}~\citeyear{2017A&A...601A..95C}, and magenta squares, \citeauthor{2018arXiv180504118G}~\citeyear{2018arXiv180504118G}). 
\label{fig:mzr}
}
\end{figure}

Figure~\ref{fig:mzr} shows the $M_\star$ versus the gas-phase metallicity relation (MZR) derived for local dwarf galaxies by \citet[][the red solid line, with the empty circles being the original data]{2012ApJ...754...98B}, as well as the relation inferred for all star-forming SDSS galaxies by \citet[][the orange solid line and shaded region]{2013ApJ...765..140A}. Both rely on the DM to infer $12+\log({\rm O/H})$. The three blue squares represent three different estimates of the UGC\,2162 metallicity as traced by the brightest knot, namely, the one inferred using HCm without [OIII]$\lambda$4363, and the two Monte-Carlo based estimates using HCm and the DM (Sect.~\ref{sec:obs}, with the values given in Table~\ref{tab:logoh}). Thus, UGC\,2162 is a high metallicity outlier of the MZR defined by local star-forming dwarfs. We have used $M_\star= 2^{+2}_{-1}\times 10^7 M_\odot$, which was inferred from the absolute magnitude and color by \citet{2017ApJ...836..191T}.

The fundamental metallicity relation  \citep[FMR;][]{2008ApJ...672L.107E,2010MNRAS.408.2115M,2010A&A...521L..53L} links the metallicity and 
the SFR of a galaxy. For the same stellar mass, objects with larger SFR are also metal poorer.  Thus, according to the FMR, the section of the MZR plane occupied by UGC\,2162 corresponds to galaxies having low SFR for their mass, which is indeed the case of UGC\,2162. Objects occupying this section of the plane appear only when the surveys are not biased toward galaxies having intense emission lines, as it has been shown by, e.g., \citet{2017A&A...601A..95C}. Low mass galaxies are faint, therefore, good metallicities are available only for those with the strongest emission lines, where the estimates are more precise. Thus, the empirical MZRs are inevitably biased toward high SFR objects in a way that depends on the galaxy luminosity, and so, on the galaxy mass.  The intermediate redshift galaxies studied by \citet[][with $0.13 <$ redshift $< 0.88$]{2017A&A...601A..95C} have been included in Fig.~\ref{fig:mzr} for illustration (the green stars). Even if these objects are emission line galaxies, their selection criteria do not include detecting [OIII]$\lambda$4363, and they present a large scatter in the $12+\log({\rm O/H})$ versus $M_\star$ plane. Some of them overlap with the region where  UGC\,2162 is located. The bias toward high SFR objects influences the current determination of the MZR, but the importance of the effect remains unclear. The impact is expected to be more pronounced at the low-mass end, and it has to be determined by comparison of the current estimates with new ones based on volume limited samples of galaxies \citep[e.g.,][]{2013A&A...554A..20S}, or MZRs corrected for Malmquist and surface brightness biases, as it is done when measuring luminosity functions \citep[e.g.,][]{2000ApJS..129....1T,2005ApJ...631..208B}.

\begin{figure}[ht!]
\includegraphics[width=10cm, bb= 70 50 560 450]{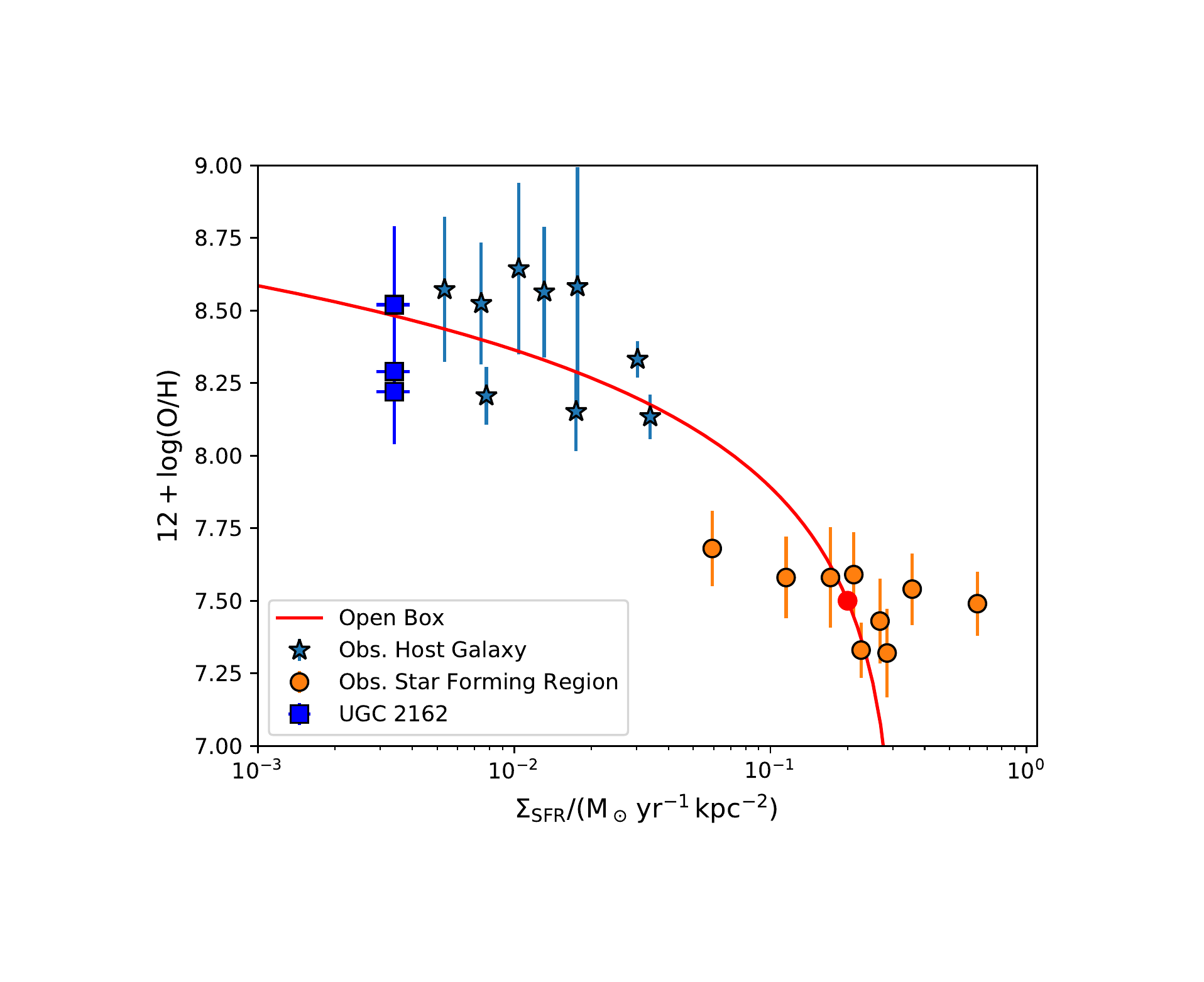}
\caption{Oxygen abundance versus surface SFR for the bright star-forming region (circles) and the host galaxy (stars) of the objects analyzed by \citet{2015ApJ...810L..15S}.  Host galaxies have high metallicity and low surface SFR compared to the starbursts. UGC\,2162 is also shown as the blue squares, and it appears in the realm of the host galaxies. (We show the same three estimates included in Fig.~\ref{fig:mzr}, with the surface SFR taken from \citeauthor{2017ApJ...836..191T}~\citeyear{2017ApJ...836..191T}.) The solid line has been included to guide the eye, and it traces the open-box evolution of the starburst shown as a red bullet, that moves from right to left along the line while SF consumes gas and produces metals \citep[details are given in ][]{2015ApJ...810L..15S}.
\label{fig:fig2}
}
\end{figure}

The gas-phase metallicity and the surface brightness of  UGC\,2162 agree with the properties expected for the host galaxies having an XMP starburst analyzed by \citet{2015ApJ...810L..15S}. Figure~\ref{fig:fig2} shows a scatter plot of the oxygen abundance versus surface SFR for the metal-poor star-forming region and the host galaxy of the ten galaxies studied by \citet{2015ApJ...810L..15S}.  Host galaxies have high metallicity and low surface SFR compared to the star-forming regions. UGC\,2162 has been represented in the Fig.~\ref{fig:fig2} including the three different metallicity estimates of the brightest knot and a common surface SFR of  $\Sigma_{\rm SFR}=3.4(\pm 0.5)\times 10^{-3} M_\odot\, {\rm yr}^{-1}\, {\rm kpc}^{-2}$. This $\Sigma_{\rm SFR}$ was inferred by \citet{2017ApJ...836..191T} from the H$\alpha$ flux of the brightest star-forming knot through the calibration of \citet{1998ARA&A..36..189K}. 
An inspection of Fig.~\ref{fig:fig2} shows that the surface SFR and the metallicity of UGC\,2162 naturally fit the values inferred for the host galaxies\footnote{UGC\,2162 is represented in Fig.~\ref{fig:fig2} through the metallicity and surface SFR of the 1st bright point. The metallicity inferred for the diffuse component of the galaxy is somewhat larger ($\sim 0.4$\,dex; Table~\ref{tab:logoh}, column\,1), but this diffuse component also has a significantly smaller $\Sigma_{\rm  SFR}$ ($\sim 1/3$, attending to the ratio of H$\alpha$ fluxes). Thus, we note that even the two components of UGC\,2162 (diffuse and bright point) seem to follow the global trend in Fig.~\ref{fig:fig2}, where metallicity increases with decreasing $\Sigma_{\rm  SFR}$. However, we are reluctant to over-interpret this result since the difference in metallicity between diffuse component and bright point may not be significant according to the error budget worked out in Sect.~\ref{subsec:metallicity}.}. Moreover, its stellar and gas masses are also typical of an XMP galaxy \citep[for reference, see][]{2013A&A...558A..18F}. Thus, UGC\,2162 has the properties to be expected  an XMP galaxy in a pre or post starburst phase, when the bright starburst has faded away and only the host galaxy remains.  Its mere existence cleans up the difficulty posed by the lack of host-like galaxies without starburst (Sect.~\ref{sec:intro}).

UGC\,2162 is a galaxy with high gas-phase metallicity for its mass. It is not unique. Some of the emission line galaxies at high redshift studied by \citet{2017A&A...601A..95C} also belong to this regime (Fig.~\ref{fig:mzr}).  In addition, \citet{2018arXiv180504118G} have recently characterized two diffuse dwarf galaxies in the local Universe similar to UGC\,2162 in this respect, namely, LSBG-285 and LSBG-750. They are low stellar mass galaxies ($M_\star\simeq  2$\,--\,$3\times 10^7\,M_\odot$) with a gas phase metallicity around 1/2 solar, which makes them high-metallicity outliers of the gas-phase MZR (the magenta squares in Fig.~\ref{fig:mzr}). These two objects have their stellar metallicity ($Z_\star$) estimated from the continuum emission. Although with large uncertainty, $Z_\star$ is estimated to be low, between 3\,\% and 10\,\% of the solar metallicity. Thus, what is expected if the stars were produced during past starbursts feeding from metal-poor gas. Other works also show high metallicity outliers of the MZR, e.g., \citet[][in the local Universe]{2008ApJ...685..904P} and \citet[][at redshift $< 0.4$]{2012ApJ...750..120Z}.


\section{Discussion and Conclusions}\label{sec:conclusions}

Using spectra from the 10-m GTC telescope, we confirm the relatively high metallicity of the ionized gas forming stars in the UDG  UGC\,2162 (Sect.~\ref{sec:obs} and Table~\ref{tab:logoh}). This result together with its low stellar mass make UGC\,2162 a high metallicity outlier of the MZR (Fig.~\ref{fig:mzr}). Its surface SFR is also low, so that  UGC\,2162 seems to have all the properties characterizing the host galaxies associated with the XMP galaxies (Sects.~\ref{sec:intro} and \ref{sec:results}). XMP galaxies have a low light-weighted mean metallicity ($Z_g < Z_\odot/10$) which, however, is often not uniform. The metallicity is low only in the young off-center bright starburst that often gives XMP galaxies their characteristic tadpole morphology (see Sect.~\ref{sec:intro}). However, the underlying galaxy hosting the starburst is significantly more metallic than the starburst. The properties of   UGC\,2162 make it a candidate to be one of such host galaxies (i.e., a {\em headless tadpole galaxy}).

Detecting galaxies with the properties of the hosts is essential for consistency, because they are expected if the lopsided starburst is created by external gas accretion. These hosts are galaxies leftover after a major star-formation phase, or the precursors of XMP galaxies before a gas accretion event triggers new star-formation episodes. We have found that UGC\,2162 may be one of them. Moreover, there are also other galaxies populating the high-metallicity region of the mass metallicity plane (see Sect.~\ref{sec:results}).  These objects are usually underrepresented in emission line galaxy surveys because they are faint, low surface brightness, and with lines of low equivalent width. However, they may represent a fundamental phase in the star-formation process of low mass galaxies, which have a bursty SF history, with periods of high SF interleaved with others of inactiveness. 

Systematic searches for XMP galaxies based on galaxy spectra provide objects that usually comply with the definition of Blue Compact Dwarf (BCD) galaxies \citep[e.g.,][]{2000A&ARv..10....1K,2011ApJ...743...77M,2016ApJ...819..110S}. Thus, the XMPs mentioned  in this paper, with luminous lopsided HII regions, seem to correspond to extreme cases of the more common BCDs. The same kind of duty cycle hypothetically linking XMPs and UDGs also fits in the relation between BCDs and their quiescent counterparts (QBCD). BCD galaxies are metal-poor systems for their stellar mass, presently going through a star-forming phase. Thus, they are relatively luminous with high surface brightness. Their outskirt light is dominated by the host galaxy, so that star-forming regions and host can be separated out \citep[e.g.,][]{2007A&A...467..541A}. Using  the properties of the host galaxies to search for QBCDs, one infers the existence of a population of QBCDs 30 times more numerous than the BCDs \citep{2008ApJ...685..194S}. As expected if BCDs and QBCDs  alternate their roles cyclicly, $Z_\star$ is the same in both types of galaxies, and agrees with $Z_g$ in BCDs, during the star-forming phase of the cycle \citep{2009ApJ...698.1497S}. On the other hand, $Z_g$ in QBCDs is significantly larger than in BCDs (0.35~dex), as if their star-forming gas was already metal enriched. Their duty cycle also concurs with the luminosity-weighted ages of their respective stellar populations, being young in BCDs ($\lesssim 1$\,Gyr) and older in QBCDs (from 1 to 10\,Gy) \citep{2009ApJ...698.1497S}.
We hypothesize that the evolution of a single galaxy might be like this: from an initially quiescent state, with $Z_g > Z_\star$ as a result of a few Gyr of stellar evolution, and with $Z_\star$ equal to that in other quiescent galaxies of the same $M_\star$,  an accretion event brings in relatively low metallicity gas with $Z_{g,in} < Z_g$, making a local star formation spot with low metallicity compared to other regions in the galaxy. This would be the phase observed as BCD (or XMP)  because of the increase in surface brightness and luminosity caused by star formation. $Z_{g,in}$ is comparable to  $Z_\star$ at the beginning of this phase, but soon after, star formation makes $Z_\star$ slightly larger than $Z_{g,in}$ and increases the stellar mass of the galaxy as well.  When star formation stops the galaxy turns quiescent again, increasing $Z_g$ over the next few Gyr. As a result, $Z_g > Z_\star$ in the new quiescent phase, with the stellar metallicity being greater than it was before, but it still consistent with the mass metallicity relation as the stellar mass has also increased.

From the point of view of the cosmological simulations of galaxy formation, low-mass high-metallicity systems are to be expected \citep[e.g.,][]{2016MNRAS.459.2632D,2017MNRAS.472.3354D,2018ApJ...859..109S}. Fixed $M_\star$, the galaxies of high SFR have star-forming gas of low metallicity. When this gas is consumed through SF and outflows, the small fraction that remains forming stars is contaminated by  metals injected during the past SF episodes, and thus, the leftover gas tends to be fairly metallic. At this point the galaxy is under-luminous and of low surface brightness. Thus,  galaxies like UGC\,2162 would represent systems in the late stages of consuming the gas that led to the last major outburst. This scenario fits in well the formation processes for UDGs found in  zoom-in cosmological simulations from the Numerical Investigation of a Hundred Astrophysical Objects (NIHAO) project \citep{2017MNRAS.466L...1D}. Some of their objects are UDGs.  They reside in isolated haloes, have $M_\star$ of $10^{7-8.5}\, M_\odot$,  effective radii larger than 1 kpc, and dark matter cores rather than cusps. They exhibit a broad range of colors, and a non-negligible $M_g$ of $10^{7 - 9}\, M_\odot$. The presence of gas turns out to be crucial to form UDGs. Gas needs to be accreted to trigger SF. Then feedback-driven gas outflows, and the subsequent dark matter and stellar expansion, are the key to reproduce faint extended galaxies, in a process similar to the creation of dark matter cores in dwarfs \citep[e.g.,][]{2010Natur.463..203G}. Somehow, dwarf galaxies with particularly bursty and prolonged SF histories tend to be extended in size, therefore, have low surface brightness when in between bursts. 
In a very recent paper, by \citet{2018MNRAS.478..906C} reach a similar conclusion, and they also predict the UDGs to have low $Z_\star$.

The need of significant amounts of gas that cycles between the galaxy and its surrounding medium seems to be a central ingredient for the creation of very low surface brightness systems. UGC\,2162 is a fairly gas rich object thus fitting well with the idea. However, for  UGC\,2162 to be truly consistent with the picture, its large gas reservoir cannot participate in the present residual SF. It has to be gas that either will fuel future SF once it settles down onto the disk, or gas driven out by past SF episodes. If the total mass of gas were participating in the SF, then according to the Kennicutt-Schmidt relation \citep[K-S relation; e.g.,][]{2012ARA&A..50..531K} the total SFR of the galaxy\footnote{Assuming $M_{\rm HI}=1.9\times 10^{9}\,M_\odot$ and a HI radius three times the effective radius \citep[i.e., 5.1\,kpc;][]{2017ApJ...836..191T}, one gets a surface gas density of $3.3\,M_\odot\,{\rm pc}^{-2}$. Then the K-S relation in \citet[][]{2012ARA&A..50..531K} gives a surface SFR of $6.6\times 10^{-4}\,M_\odot\,{\rm yr^{-1}}\,{\rm kpc}^{-2}$, which integrated over the optical galaxy gives a SFR of $6.0\times 10^{-2}\,M_\odot\,{\rm yr}^{-1}$.} would be around $6\times 10^{-2}\,M_\odot\,{\rm yr}^{-1}$. The SFR of the main SF knot in UGC\,2162 is around $8.7\times 10^{-5}\,M_\odot\,{\rm yr}^{-1}$ \citep{2017ApJ...836..191T}, so that even if tens of knots like this one contribute to the total SF, they will never sum up to give the SFR expected if all the observed HI mass participates in the SF process.    

\acknowledgements

The work has been partly funded by the Spanish Ministry of Economy and  Competitiveness (MINECO), project {\em Estallidos}  AYA2016-79724-C4-2-P.  
M.\,E.\,F. gratefully acknowledges the financial support of the {\em Funda\c{c}\~ao para a Ci\^encia e Tecnologia} (FCT--Portugal), through the grant SFRH/BPD/107801/2015.
%
%
Thanks are due to an anonymous referee for comments that led to clarifying the presentation of results.
Based on observations made with the Gran Telescopio Canarias (GTC), instaled in the Spanish Observatorio del Roque de los Muchachos of the Instituto de Astrof\'\i sica de Canarias, in the island of La Palma.
\medskip
\software{Python, TOPCAT}.

\vskip 4cm

\bibliography{ms}
\bibliographystyle{aasjournal}



\end{document}